\documentstyle[12pt]{article}  \hoffset
-1.0cm \voffset -0.5cm \textheight 24.0cm \textwidth 16.5cm
\begin{document}
%%%%%%%%%%%%%%%%%%%%%%%%%%%%%%%%%%%%%%%%%%%%%%%%%%%%%%%%%%%%%%%%%%%%%%%%%%%%
\def\theequation{\arabic{section}.\arabic{equation}}
%%%%%%%%%%%%%%%%%%%%%%%%%%%%%%%%%%%%%%%%%%%%%%%%%%%%%%%%%%%%%%%%%%%%%%%%%%%%%
\title {TENSOR-TENSOR THEORY OF GRAVITATION} \author{\bf {Merab
Gogberashvili} \\ Inst. of Physics, Tamarashvili str. 6, 380077 Tbilisi,
Republic of Georgia \\ E-mail: gogber@physics.iberiapac.ge} \maketitle
%%%%%%%%%%%%%%%%%%%%%%%%%%%%%%%%%%%%%%%%%%%%%%%%%%%%%%%%%%%%%%%%%%%%%%%%%%
\begin{abstract} We consider the standard gauge theory of Poincar\'{e}
group, realizing as a subgroup of $GL(5. R)$. The main problem of this
theory was appearing of the fields connected with non-Lorentz symmetries,
whose physical sense was unclear. In this paper we treat the gravitation as a
Higgs-Goldstone field, and the translation gauge field as a new tensor
field. The effective metric tensor in this case is hybrid of  two
tensor fields. 

In the linear approximation the massive translation gauge field can give
the Yukava type correction to the Newtons potential. Also outer potentials
of a sphere and ball of the same mass are different in this case. 

Corrections to the standard Einshtein post Newtonian formulas of the light
deflection and radar echo delay is obtained. 

The string like solution of the nonlinear equations of the translation
gauge fields is found. This objects can results a Aharonov-Bohm type effect
even for the spinless particles. They can provide density fluctuations in
the early universe, necessary for  galaxy formations. 

The spherically symmetric solution of the theory is found. The translation
gauge field lead to existence of a impenetrable for the matter singular
surface inside the Schwarzschild sphere, which can prevent gravitational
collapse of a massive body. \end{abstract}

\vskip 1cm
 

%%%%%%%%%%%%%%%%%%%%%%%%%%%%%%%%%%%%%%%%%%%%%%%%%%%%%%%%%%%%%%%%%%%%%%%%%%%
\newpage 
\tableofcontents
%%%%%%%%%%%%%%%%%%%%%%%%%%%%%%%%%%%%%%%%%%%%%%%%%%%%%%%%%%%%%%%%%%%%%%%%%%%%%%%%%%
\newpage
\section{Introduction}

~~~~~Most variational principles of current interest in physics are
manifestly invariant under the ten-parameter Poincar\'{e} group $$ P_{10}
= L_{6} \rhd T_{4} , $$ where $L_{6}$ is the six-parameter Lorentz group,
$T_{4}$ is the four-parameter translation group, and $\rhd$ denotes the
semidirect product.  In addition $P_{10}$ is the maximal group of
isometries of the Minkowski space-time. Soon after establishing of the
gauge formalism for the groups of internal symmetries gauge theories based
on the $P_{10}$ appear (see \cite{P} and references therein).  The
intrinsic difficulties in gauging the $P_{10}$ were arises immediately.
The reason is that $P_{10}$ is not a semisimple group, and because it acts
both on the matter fields and on the underlying space-time manifold. 

The gauge theories has most elegant form in the fibre bundle formalism
\cite{DM}.  In this formalism two kinds of the gauge transformations
generally are considered: 

1. Passive gauge transformations which remain the matter
  fields constant, but change atlas of the fibre.

2. Active gauge transformations, which transform the matter fields themself. 

The standard gauge formalism and the principle of the local invariance is
formulated for the gauge transformations of the first kind.  In case of
the internal symmetries to the gauge transformations of the first kind
correspond gauge transformations of the second kind with the same matrix
form. So for the Yang-Millse fields gauge transformations usually are not
differentiated. 

As was mentioned above, difficulties with the gauge transformations arises
in the case of space-time transformations, which acts on the derivative
operators $\partial_{\nu}$, as well as on the vectors of the tangent space
on a space-time manifold.  Here we have two types of the gauge
transformation of the first kind. 

a). Transformation of a  atlas of the only tangent bundle.

 The principle of relativity is formulated as the requirement of symmetry
of the atlas of the tangent bundle with respect to the general linear
group. The principle of equivalence reduces this group to the $L_{6}$.
This invariance of the theory besides of gauge fields - Lorentz connection
$\Gamma_{\nu}$, is provided by introduction of the metric field
$g_{\mu\nu}$, or isomorphic to him tetrad field $h^{A}_{\mu}$ . The gauge
transformation of this kind act on $h^{A}_{\mu}$ from the left,
transforming index $\mu$, which corresponds to the atlas of the tangent
bundle.  They also act on $\Gamma_{\nu}$, and on the indexes of the
covariant derivative operators. The invariance of the Lagrangian leads to
the conservation law for the energy-momentum tensor. This kind of
transformations can be connected with the gauge theory of gravity, but
still remains open the question about gauge status of the gravitational
field $h^{A}_{\mu}$. 

In this paper we consider the theory connected to the second type of 
the gauge transformations of the first kind:

 b). Transformation of a atlas of the only matter fields. 

Invariance of the theory is provided by introduction of the only gauge
fields. Bundle of the matter fields is associated with the tangent bundle,
so the principle of equivalence contracts it structural group too. For the
correlation of the symmetries of both bundles the structural group of
matter fields bundle reduces to the $P_{10}$ immediately, in difference
with the tangent bundle, where one must introduce the metric tensor
$g_{\mu\nu}$. So, except of the gauge fields of $L_{6}$, this theory
contains gauge fields of translations $\theta^{\nu}_{\mu}$. Gauge
transformations of this type acted on matter fields, on the tetrad
functions $h^{A}_{\mu}$ from the right, changing index $A$, which
corresponded to the atlas were fields are defined, and on $\Gamma_{\nu}$.
They provides conservation laws, usual for gauge theories. For this type
of symmetries the physical mining of gauge fields $\theta^{\nu}_{\mu}$ was
unclear. 

Because of importance of the question we claim again that tetrad
gravitational fields $h^{A}_{\mu}$ and gauge field of translations
$\theta^{\nu}_{\mu}$ can not be identified \cite{S, SG}. Main difference
is that $\theta^{\nu}_{\mu}$ has both coordinate indexes and $h^{A}_{\mu}$
has one coordinate and one tetrad indexes. This results different
transformation laws of this fields under transformations of space-time
groups. Also $h^{A}_{\mu}$ can be zero, but $\theta^{\nu}_{\mu}$ not; 
$\theta^{\nu}_{\mu}$ is single valued, $h^{A}_{\mu}$ - with accuracy of
Lorentz transformations and so on. 

The conventional gauge technique can be applied to the $P_{10}$ if one
ignore its physical role as a dynamical group and looks at it as an
abstract structural group of a bundle, providing gauge transformations of
first kind, of type b).  In this paper we follow the formalism of direct
gauging of the abstract Poincar\'{e} group \cite{Ed}. In difference with
this works gauge fields of $P_{10}$ we don't connect with the gravity, but
treat as a new fields, describing new fundamental interaction in the
space-time. 

The gravity in our scheme appear not as the gauge, but Higgs-Goldstone
type field \cite{S}, usually considered in gauge theories. In the gauge
gravitational theory the equivalence principle leads to contraction of the
structural group of the tangent bundle to the $P_{10}$, imitating the
situation, which is analogous to the spontaneous symmetry breaking.  Then
we can look upon the Minkowski metric field as being the vacuum Higgs
field, and small perturbations may play role of Goldstone fields.  This
metric perturbations can be identified with the presence of a
gravitational field \cite{S}. 

So our approach is hybrid of the two gravitational theories. The effective
metric tensor in our case depends on two tensor fields - gravitational
field and Poincar\'{e} gauge fields. So we generalize known scalar-tensor
and vector-tensor theories of the gravitation (see references in
\cite{Wi}). 

In Section 2 we present a brief review in gauge theory of dislocations 
and disclinations in a elastic medium. Section 3 develops the formalism of 
the gauge theory of the group $GL(5, R)$ and general form of equations of 
Poincar\'{e} gauge fields are obtained. Then we consider only the case 
when Lorentz gauge fields are zero. In Section 4 we express the theory of 
translation gauge field. In rest Sections 5, 6, 7 and 8 the solutions 
of translation gauge field equations and possible effects of this field are 
considered.

%%%%%%%%%%%%%%%%%%%%%%%%%%%%%%%%%%%%%%%%%%%%%%%%%%%%%%%%%%%%%%%%%%%%%%%%%%%%%%%%
\newpage
\section {The gauge theory of dislocations and disclinations}
\setcounter{equation}{0}

~~~~~The question about the physical sense of gauge fields corresponding
to 3-rotations and 3-translations had been discussed also in the gauge
theory of dislocations and disclinations for the elastic medium \cite{KE}. 
Here we remind the main basis of this theory. 

The gauge theory of dislocations and disclinations is based on the fact
that diffeomorphism
\begin{equation}
\chi^{i} = u^{i} + \delta^{i}_{j}x^{j} , ~~~~~i,j = 1, 2, 3,
\label{2.1}
\end{equation}
which characterizing displacement $u^{i}$ of the point $x^{i}$  of the 
elastic
medium in the case of a small deformation is determined only with 
the accuracy to local gauge transformations
\begin{equation}
\chi^{i} \longrightarrow \chi^{i} + \varepsilon^{i}_{kl}\chi^{k}A^{l}(x)
\ + B^{i}(x)  ,
\label{2.2}
\end{equation}
where $A^{i}(x)$ and $B^{i}(x)$ are local rotations and
local translations.  Corresponding gauge fields $W^{i}_{j}$ and
$\Theta^{i}_{j}$ are forming plastic part $\beta^{P}_{ij}$ of total
distorsion
\begin{equation}
\beta^{T}_{ij} = \partial_{j}\chi_{i} = \beta_{ij} + \
\beta^{P}_{ij} .
\label{2.3}
\end{equation}

The "covariant" derivative
\begin{equation}
D_{j}\chi_{i} = 
\beta^{T}_{ij} - \beta^{P}_{ij} = \partial_{j}\chi_{i} - 
\varepsilon_{ikl}\chi^{k}W^{l}_{j} - \Theta_{ij} ,
\label{2.4}
\end{equation}
coincides with the elastic distorsion $\beta_{ij}$. Gauge fields strengths
\begin{eqnarray}
F^{i}_{kl} = \partial_{k}W^{i}_{l} - \partial_{l}W^{i}_{k} + 
C^{i}_{mn}W^{m}_{k}W^{n}_{l}  ,~~~~~~~~  \nonumber \\
\alpha^{i}_{kl} = \partial_{k}\Theta^{i}_{l} - 
\partial_{l}\Theta^{i}_{k} + \varepsilon^{i}_{mn}(W^{m}_{k}\Theta^{n}_{l} 
 - W^{m}_{l}\Theta^{n}_{k} + F^{m}_{kl}\chi^{n})  ,
\label{2.5}
\end{eqnarray}
are densities of dislocations and disclinations respectively. In this
case
$\beta^{i}_{j}$ ,
$F^{i}_{kl}$
and
$\alpha^{i}_{kl}$
appear to be gauge invariant quantities.

The Lagrangian of the gauge theory of continuous defects is choose in the
 following form:
\begin{equation}
L = \frac{1}{4}(2\mu e^{i}_{j}e^{j}_{i} + 
\lambda e^{i}_{i}e^{j}_{j}) - \frac{a}{2}\alpha^{i}_{kl}\alpha^{kl}_{i} - 
\frac{b}{2}F^{i}_{kl}F^{kl}_{i}  ,
\label{2.6}
\end{equation}
 where 
\begin{equation}
e_{ij} = D_{i}\chi^{k}D_{j}\chi_{k} - \delta_{ij}
\label{2.7}
\end{equation}
is nonlinear tensor of deformations.  In (\ref{2.6}) coupling constants
$\mu$ and $\lambda$ corresponded to Lam\'{e} parameters, describing
elastical properties of the media. Parameters $a$ and $b$ characterizing
the energies required in order to create "unit"  dislocation and
disclination, respectively. 

Equations of theory has the following general structure:
\begin{eqnarray}
(\delta^{i}_{m}\partial_{j} - 
\varepsilon^{i}_{km}W^{k}_{j})\frac{\partial L}{\partial(D_{j}\chi^{i})} = 
\varepsilon^{i}_{km}F^{k}_{jl}\frac{\partial L}{\partial\alpha^{i}_{jl}} , 
\label{2.8} \\
(\delta^{i}_{m}\partial_{j} -                                           
\varepsilon^{i}_{km}W^{k}_{j})\frac{\partial L}{\partial\alpha^{i}_{jl}} = 
 - \frac{1}{2}\frac{\partial L}{\partial(D_{l}\chi^{m})} ,
\label{2.9} \\
(\delta^{i}_{m}\partial_{j} - 
C^{i}_{km}W^{k}_{j})\frac{\partial L}{\partial F^{i}_{jl}} = 
\varepsilon^{i}_{mj}(\frac{\partial L}{\partial(D_{l}\chi^{i})}\chi^{i} +
2\frac{\partial L}{\partial\alpha^{i}_{lk}}\Theta^{j}_{k}) .
\label{2.10}
\end{eqnarray}

From the experiments is known, that disclination energy is large compared
with the dislocation energy and this is very large compared with the
elastic energy. So, $\lambda/a \sim a/b \ll 1$. Then in the first order of
this scaling parameters from (\ref{2.8}) we get well known equation of
equilibrium of the continuous media 
\begin{equation}
\partial_{j}\frac{\partial\overline{L}}{\partial(D_{j}\chi^{i})} =
\partial_{j}\overline{\sigma^{j}_{i}} = 0  ,
\label{2.11}
\end{equation}
where strength tensor
$\overline{\sigma^{j}_{i}}$
is expressed by elastic deformation (in the linear approximation)
\begin{equation}
\overline e_{ij} = D_{i}u_{j} + D_{j}u_{i} ,
\label{2.12}
\end{equation}
 in correspondence of the Hooke's law for the isotropic media:
\begin{equation}
\overline{\sigma^{j}_{i}} = \mu\overline e^{j}_{i} +
\frac{\lambda}{2}\delta^{j}_{i}\overline e^{k}_{k} .
\label{2.13}
\end{equation}
Lam\'{e} coefficients $\mu$ and $\lambda$ are positive and provided
positiveness of the elastic energy.

Equations (\ref{2.9}) and (\ref{2.10}) are equations for dislocations and
disclinations respectively.  The source of dislocations $\Theta^{i}_{j}$ is
nonlinear tensor of strength
\begin{equation}
\sigma^{i}_{j} = \frac{\partial L}{\partial (D_{j}\chi^{i})}
\label{2.14}
\end{equation}
and the source of disclination, besides of $ \sigma^{i}_{j} $ is the torsion
of media also. 

Dislocations appear in the second order approximation and only in the
third and higher order approximations do the disclination field
$W^{i}_{j}$ enter the equations. So in case of a weak deformation of the
media disclinations are not created. 

%%%%%%%%%%%%%%%%%%%%%%%%%%%%%%%%%%%%%%%%%%%%%%%%%%%%%%%%%%%%%%%%%%%%%%%%%%%%
\newpage
\section{ General theory}
\setcounter{equation}{0}

~~~~~Success of the gauge theory of continuous defects \cite{KE}, where a
tangent vector corresponds to the vector of medium displacement and gauge
fields of 3-translations and 3-rotations interpreted as dislocations and
disclinations (linear defects of a crystal lattice), allows us to
interpret gauge fields of the nondynamical Poincar\'{e} group as the
"defects" of the space-time manifold \cite{SG}. The gravity appeared in
the our scheme as the Higgs-Goldstone field, as in papers \cite{S}. 

The standard Yang-Millse gauge theory of the group $P_{10}$
can be constructed by presenting of $P_{10}$ as a matrix subgroup of 
$GL(5,R)$  \cite{Ed} 
\begin{equation}
P_{10} = \left(
\begin{array}{cc}
 L_{6} & T_{4}   \\
 0     & 1
\end{array}
\right) ,
\label{3.1}
\end{equation}
where $L_{6}$ and $T_{4}$ denote transformation matrices of the Lorentz group
and translation group respectively. This matrices induce
diffeomorphisms of the five dimensional space - gauge transformations of the
second kind:
\begin{equation}
x \longrightarrow P_{10}x = \
\left\{ \begin{array}{c}
L_{6}x + T_{4} \\ 1
\end{array} \right\} .
\label{3.2}
\end{equation}

Let $I_{A}$ ($A = 0, \ldots , 5 $) be the basis of the algebra of group
$L_{6}$ and $e_{\mu}$ ($ \mu = 0, \ldots , 3$) be the generators of
group $T_{4}$.  Then the standard Yung-Millse connection has the form:
\begin{equation}
\Gamma = \left(
\begin{array}{cc}
W^{A}I_{A} & \theta^{\mu}e_{\mu}  \\
0          & 0 
\end{array}
\right) ,
\label{3.3}
\end{equation}
where $W^{A}_{\mu}$ and $\theta^{\nu}_{\mu}$ are gauge fields of groups
$L_{6}$ and $T_{4}$ respectively. Covariant derivatives and 
differentials can be written in the following form:
\begin{eqnarray}
\tilde{\partial_{\mu}} = (h^{-1})^{\alpha}_{\mu}\partial_{\alpha} , 
\nonumber \\
\tilde{d}x^{\mu} = h^{\mu}_{\nu}dx^{\nu} ,  \label{3.4}
\end{eqnarray}
where
\begin{equation}
h^{\mu}_{\nu} = \delta^{\mu}_{\nu} + I^{\mu}_{A\gamma}x^{\gamma}W^{A}_{\nu} \
+ \theta^{\mu}_{\nu}
\label{3.5}
\end{equation}
denotes distorsion for Poincar\'{e} group, and $(h^{-1})^{\alpha}_{\mu}$
is it reverse matrix. Then  elements of length and volume in the 
Minkowski space  can be transform to:
\begin{eqnarray}
ds^{2}\longrightarrow\tilde{d}s^{2} = \eta_{\mu\nu}h^{\mu}_{\alpha} 
h^{\nu}_{\beta}dx^{\alpha}dx^{\beta} = 
\tilde{g_{\alpha\beta}}dx^{\alpha}dx^{\beta} ,  \nonumber \\
dV \longrightarrow \tilde{d}V = hdV ,~~~~~~~~~~ \label{3.6}
\end{eqnarray}
where $h$  is the determinant of $h^{\mu}_{\nu}$. In this case arbitrary 
4-vector     is transforming as a 4-vector of velocity
\begin{eqnarray}
\tilde{v}^{\mu} = \frac{dx^{\mu}}{\tilde{d}s} = h^{\mu}_{\nu}v^{\nu} , 
\nonumber \\
\tilde{v}_{\mu} = (h^{-1})^{\nu}_{\mu}v_{\nu} , \label{3.7}
\end{eqnarray}
i. e. in the form:
\begin{eqnarray}
\tilde{B}^{\mu} = h^{\mu}_{\nu}B^{\nu} , 
\nonumber \\
\tilde{B}_{\mu} = (h^{-1})^{\nu}_{\mu}B_{\nu} , \label{3.8}
\end{eqnarray}

By the connection (\ref{3.3})  one can build up a Yung-Millse 
curvature for this case
\begin{equation}
R = d\Gamma + \Gamma \wedge \Gamma .
\label{3.9}
\end{equation}
Then we can obtain curvatures corresponding to the $L_{6}$ and $T_{4}$ groups
respectively:
\begin{eqnarray}
F^{A} = dW^{A} + \frac{1}{2}C^{A}_{BC}W^{B} \wedge W^{C}  , \nonumber \\
\alpha^{\mu} = d\theta^{\mu} + C^{\mu}_{A\nu}W^{A}\wedge\theta^{\nu} .
\label{3.10}
\end{eqnarray}
 Here we have used  properties of structural constants
\begin{eqnarray}
C^{\mu}_{AB} = C^{\nu}_{\alpha\beta} = C^{A}_{\alpha\beta} = 0 ,
\nonumber \\
(A,B, ... = 0,...,5 , \mu, \alpha, \beta, ... = 0,...,3)
\label{3.11}
\end{eqnarray}
of the Poincar\'{e}  group.

In the ordinary way we can get also the expression of the torsion tensor
\begin{equation}
S^{\mu} = \alpha^{\mu} + F^{A}I^{\mu}_{A\nu}x^{\nu} .
\label{3.12}
\end{equation}

So localization of the  nondinamical Poincar\'{e} group lifts Minkowski 
space into the 
nonholonomic space and now in the every point we have two spaces, as in the
bimetric theories of gravitation (see references in\cite{Wi}). But we 
assume, that the effective metric 
\begin{equation} 
\tilde{g}_{\mu\nu} =
\eta_{\alpha\beta}h^{\alpha}_{\mu}h^{\beta}_{\nu} \label{3.13}
\end{equation} 
is not contain the gravitational field. It constructed by means of 
the distorsion
(\ref{3.5}), both indexes of which are coordinate and not with the tetrad
gravitational field $h^{A}_{\mu}$. So our theory differs from other gauge
models of the Poincar\'{e} group \cite{P} by the fact that in our case 
the 
distorsion $h^{\mu}_{\nu}$ and torsion $S^{\mu}_{\alpha\beta}$ depends on
the translation gauge fields, as well as on the Lorentz gauge fields 
$W^{A}_{\mu}$,
and differs from the model \cite{Ed}, where distorsion (\ref{3.5}) is
treated as the gravitational field.  Gauge theory of $GL(5,R)$ considered in
this section must be supplied various kind of Goldstone and Higgs field
appearing, which has not place in scheme of theory \cite{Ed}. In our model
this fields connected to the gravitational field $g_{\mu\nu}$. In the
presence of gravitational field background Minkowski space becomes the
Riemann space.  Then the effective metric tensor and effective connection in
this case -
\begin{eqnarray}
\tilde{g}_{\mu\nu} = g_{\alpha\beta}h^{\alpha}_{\mu}h^{\beta}_{\nu} = 
\eta_{AB}H^{A}_{\mu}H^{B}_{\nu} ,  \label{3.14} \\
\tilde{\Gamma}^{\alpha}_{\mu\nu} = h^{\gamma}_{\mu}h^{\delta}_{\nu} \
(h^{-1})^{\alpha}_{\beta}\Gamma^{\beta}_{\gamma\delta} + \
(h{-1})^{\alpha}_{\beta}\partial_{\mu}h^{\beta}_{\nu} ,
\label{3.15}
\end{eqnarray}
where $\Gamma^{\beta}_{\gamma\delta}$ is ordinary Christoffels symbols,
appear to be hybrid of two fields - classical gravitation field and
Poincar\'{e} gauge fields. Thus our theory appears to be tensor-tensor and
modifies known scalar-tensor and vector-tensor theories of gravity 
(see references in \cite{Wi}).

A Poincar\'{e} gauge fields is inserted into the Lagrangian of matter fields
only via the effective metric (\ref{3.14}) and the effective connection 
(\ref{3.15}). 

Let us consider that total Lagrangian $L_{TOT}$ of matter fields
$\varphi^{a}$ and Poincar\'{e} gauge fields $\theta^{\mu}_{\nu} , 
W^{A}_{\nu}$ has the standard Yang-Millse structure
\begin{equation}
L_{TOT} = hL_{mat}(\varphi^{a}, \tilde{\partial}_{\mu}\varphi^{a}) +
hL_{\theta, W}(\theta^{\mu}_{\nu}, W^{A}_{\nu}, S^{\mu}_{\alpha\beta},
F^{A}_{\alpha\beta}) .
\label{3.16}
\end{equation}
We put for the convenience, that the  gauge fields Lagrangian instead of
$\alpha^{\mu}_{\alpha\beta}$
depends on the torsion tensor $S^{\mu}_{\alpha\beta}$, which is expressed 
by   $\alpha^{\mu}_{\alpha\beta}$ and $F^{A}_{\alpha\beta}$
by the formula (\ref{3.12}).

By variation with respect of fields $\varphi^{a}$ and $\theta^{\mu}_{\nu}$
we obtain  field equations
\begin{eqnarray}
(\delta^{b}_{a}\partial_{\mu} - W^{A}_{\mu}M^{b}_{Aa})h(h^{-1})^{\mu}_{\nu} 
\frac{\partial L_{mat}}{\partial(\partial_{\nu}\varphi^{b})} = 
h\frac{\partial L_{mat}}{\partial\varphi^{a}} ,~~~~~~~~~~~~  \label{3.17} \\
2(\delta^{\beta}_{\alpha}\partial_{\mu} - W^{A}_{\mu}C^{\beta}_{A\alpha}) 
\frac{\partial (hL_{\theta, W})}{\partial S^{\beta}_{\mu\nu}} = 
- h(h^{-1})^{\nu}_{\beta}T^{\beta}_{\alpha} + \frac{\partial (hL_{\theta, 
W})}  {\partial\theta^{\alpha}_{\nu}}\mid_{S, F} ,~~~~~  \label{3.18} \\
2(\delta^{B}_{A}\partial_{\mu} - W^{C}_{\mu}C^{B}_{CA})  
\frac{\partial (hL_{\theta, W})}{\partial F^{\beta}_{\mu\nu}} = 
 h(h^{-1})^{\nu}_{\mu}M^{a}_{Ab}\frac{\partial L_{mat}} 
{\partial(\partial_{\mu}\varphi^{a})}  - 
h^{\alpha}_{\beta}I^{\mu}_{A\alpha}\frac{\partial (hL_{\theta, W})} 
{\partial S^{\mu}_{\beta\nu}} ,  \label{3.19}
\end{eqnarray}
where $M^{b}_{Aa}$ are matrices of representation of  fields
$\varphi^{a}$ induced by infinitesimal transformations of the Poincar\'{e} 
group,
and $T^{\alpha}_{\mu}$ is the canonical energy-momentum tensor of matter
fields $\varphi^{a}$. From equations (\ref{3.18}) we see that  sources
of the translation gauge field $\theta^{\mu}_{\nu}$ are the matter
energy-momentum tensor (first term on the right side) and energy-momentum
tensor of the field $h^{\mu}_{\nu}$ itself (second term on the right side). 
From
(\ref{3.19}) we see, that sources of fields $W^{A}_{\nu}$ are the spin
tensor of the matter (first term on right side) and the tensor of spin of the
field $h^{\mu}_{\nu}$ (second term on the right side). 

%%%%%%%%%%%%%%%%%%%%%%%%%%%%%%%%%%%%%%%%%%%%%%%%%%%%%%%%%%%%%%%%%%%%%%%%%%%
\newpage
\section{Translation gauge fields}
\setcounter{equation}{0}

~~~~~In the previous section the general form of equations of Poincar\'{e}
gauge fields was constructed. Now we need to choose the Lagrangian of the
theory. In the elastic media disclinations appear only in the case of
strong deformations (see the Section 2), so it is reasonable to consider
the case when the Lorentz gauge fields are zero
\begin{equation}
W^{\mu}_{\nu} = 0 .
\label{4.1}
\end{equation}
The distorsion (\ref{3.5}) and the torsion (\ref{3.12}) tensors  now has the 
form: \begin{eqnarray}
h^{\mu}_{\nu} = \delta^{\mu}_{\nu} + \theta^{\mu}_{\nu} , 
\nonumber \\
S^{\alpha}_{\mu\nu} = \alpha^{\alpha}_{\mu\nu} = 
\tilde{\partial}_{\mu}\theta^{\alpha}_{\nu} - 
\tilde{\partial}_{\nu}\theta^{\alpha}_{\mu} .\label{4.2}
\end{eqnarray}

By analogy with the Lagrangian of the dislocation gauge theory 
(\ref{2.6}) the
Lagrangian of the gauge translation field $\theta^{\mu}_{\nu}$ is proposed
to consist the Lagrangian of displacement field $L_{u}$, under the gauge
$u = 0$, and the Lagrangian $L_{\theta}$ of the field $\theta^{\mu}_{\nu}$
itself. Since, under the gauge $u^{\nu} = 0$,
\begin{equation}
D_{\mu}u^{\nu} = - \theta^{\nu}_{\mu} ,
\label{4.3}
\end{equation}
 the Lagrangian $L_{u}$  is reduced to algebraic combinations of fields
$\theta^{\mu}_{\nu}$. By analogy with the Lagrangian (\ref{2.6}) we choose 
$L_{u}$ in the form \cite{SG}:
\begin{equation}
 L_{u} = \frac{\mu}{2}e_{\mu\nu}e^{\mu\nu} + 
\frac{\lambda}{4}e^{\alpha}_{\alpha}e^{\beta}_{\beta} ,
\label{4.4}
\end{equation}
where
\begin{equation}
e_{\mu\nu} = h^{\alpha}_{\mu}h_{\nu\alpha} - \eta_{\mu\nu} .
\label{4.5}
\end{equation}

The Lagrangian $L_{\theta}$ represents all possible combinations from
components of the strength tensor $S^{\alpha}_{\mu} $, and we choose it in
the form \cite{SG}: 
\begin{equation}
L_{\theta} = - \{\frac{1}{2a}S^{\mu}_{\nu\mu}S^{\nu\alpha}_{\alpha} + 
\frac{1}{2b}S_{\mu\nu\alpha}S^{\nu\mu\alpha} + 
\frac{1}{2c}S_{\mu\nu\alpha}S^{\mu\nu\alpha} + 
d\varepsilon^{\mu\nu\alpha\beta}S^{\gamma}_{\mu\gamma}S_{\nu\alpha\beta}\} ,
\label{4.6}
\end{equation}
Here $a, b, c, d$ are coupling constants.

To obtain constraints on the constants in the Lagrangian (\ref{4.6}), now we 
consider the linear approximation in $\theta^{\mu}_{\nu}$ and assume the 
matter to be spinless.  Then, variation of the total Lagrangian
\begin{equation}
L = L_{matt} + L_{u} + L_{\theta}
\label{4.7}
\end{equation}
over $\theta^{\mu}_{\nu}$  results in the equations
\begin{eqnarray}
\frac{1}{a}(\partial_{\mu}S^{\gamma}_{\gamma\nu} - 
\eta_{\mu\nu}\partial^{\varepsilon}S^{\gamma}_{\gamma\varepsilon}) + 
\frac{1}{b}(\partial^{\varepsilon}S_{\varepsilon\mu\nu} - 
\partial^{\varepsilon}S_{\nu\mu\varepsilon}) + 
\frac{2}{c}\partial^{\varepsilon}S_{\mu\varepsilon\nu} +  \nonumber \\
+ 2d(\varepsilon_{\varrho\sigma\gamma\tau}\eta_{\mu\nu}\partial^{\varrho} 
S^{\tau\sigma\gamma} - 
\varepsilon_{\nu\sigma\gamma\tau}\partial_{\mu}S^{\tau\sigma\gamma}) + 
2\mu\theta_{\mu\nu} + \lambda\eta_{\mu\nu}\theta^{\gamma}_{\gamma} = 
- T_{\mu\nu} , \label{4.8}
\end{eqnarray}
where $T_{\mu\nu}$ is the energy-momentum tensor of the matter fields.

The divergence of this equations with respect to the index $\nu$  takes 
on the form:
\begin{equation}
2\mu\partial^{\nu}\theta_{\mu\nu}  +
\lambda\partial_{\mu}\theta^{\gamma}_{\gamma} = 0 . \label{4.9}
\end{equation}

It seems naturally to require that the divergence of the equation 
(\ref{4.8})   
with respect to the second index $\mu$ in the spinless case takes on the 
same form as equation (\ref{4.9}) -
\begin{equation}
2\mu\partial^{\mu}\theta_{\mu\nu}  +
\lambda\partial_{\nu}\theta^{\gamma}_{\gamma} = 0 . \label{4.10}
\end{equation}

This requirement imposes the following constraints
\begin{equation}
d = 0 ,~~~~~
\frac{1}{a} + \frac{1}{b} + \frac{2}{c} = 0  \label{4.11}
\end{equation}
on the constants in the Lagrangian $L_{\theta}$.

With the constraints (\ref{4.11}) nonlinear equations of translation 
gauge fields (\ref{3.18}) has the form:
\begin{eqnarray}
\frac{1}{a}(\partial_{\mu}S^{\gamma}_{\gamma\nu} -
\eta_{\mu\nu}\partial^{\varepsilon}S^{\gamma}_{\gamma\varepsilon}) +
\frac{1}{b}(\partial^{\varepsilon}S_{\varepsilon\mu\nu} -
\partial^{\varepsilon}S_{\nu\mu\varepsilon}) -
(\frac{1}{a} + \frac{1}{b})\partial^{\varepsilon}S_{\mu\varepsilon\nu} + 
\nonumber \\
+ 2\mu e_{\gamma\mu}h^{\gamma}_{\nu} + \lambda 
e^{\gamma}_{\gamma}h_{\mu\nu} + \eta_{\mu\nu}L_{\theta, W} = - 
(h^{-1})_{\nu\beta}T^{\beta}_{\mu} . \label{4.12}
\end{eqnarray}

The equation (\ref{4.9}) has the form of the gauge condition. It's
analogous to the equilibrium equation of the elastic media (\ref{2.11}) .
This equation can be obtained from the Lagrangian by variation with
respect of the field $u^{\nu}$, which is a "displacement" in the bundle.
Similar to (\ref{4.9}) conditions are considered in bimetric theories of
gravitation \cite{Wi, Lo}, but without any connections to the variational
principle. 

If we take
\begin{equation}
\lambda = - \mu , \label{4.13}
\end{equation}
then  in the linear 
approximation, one can derive the equations for the symmetrical part,
\begin{equation}
e_{\mu\nu} = \theta_{\mu\nu} + \theta_{\nu\mu}
\label{4.14}
\end{equation}
of the free translation gauge field in the form:
\begin{eqnarray}
(\Box + m^{2})(e^{\alpha}_{\beta} -
\frac{1}{2}\delta^{\alpha}_{\beta}e^{\nu}_{\nu}) = 0 , \nonumber \\
\partial_{\alpha}(e^{\alpha}_{\beta} -
\frac{1}{2}\delta^{\alpha}_{\beta}e^{\nu}_{\nu}) = 0 . \label{4.15}
\end{eqnarray}

This equations describe fields with the mass
\begin{equation}
m^{2} = \frac{a\lambda}{2} = - \frac{a\mu}{2} ,
\label{4.16}
\end{equation}
and with the spin 2 and 0, as in the theory \cite{Lo}.

%%%%%%%%%%%%%%%%%%%%%%%%%%%%%%%%%%%%%%%%%%%%%%%%%%%%%%%%%%%%%%%%%%%%%%%%%%%%
\newpage
\section{Newton's approximation}
\setcounter{equation}{0}

~~~~~In the case of joint action of the gravitational field $g_{\mu\nu}$ and
the translation gauge field $\theta^{\mu}_{\nu}$ effective curvature tensor
has the form: 
\begin{equation}
\tilde{R}^{\alpha}_{\beta\mu\nu} 
 = h^{\gamma}_{\mu}(h^{-1})^{\alpha}_{\delta}R^{\delta}_{\beta\gamma\nu} ,
\label{5.1}
\end{equation}
where $R^{\delta}_{\beta\gamma\nu}$ is the ordinary Riemann tensor.
Then the modified Hilbert Lagrangian has the form:
\begin{equation}
L_{g} = - h\sqrt{-g}R .
\label{5.2}
\end{equation}

In the linear approximation, when
\begin{equation}
g_{\mu\nu} = \eta_{\mu\nu} + \Phi_{\mu\nu} ,
\label{5.3}
\end{equation}
and in the case of the spinless source, equations of the gravitational field 
and translation gauge field are: 
\begin{eqnarray}
\Box\Phi^{\alpha}_{\beta} - 
\partial_{\beta}\partial^{\nu}\Phi_{\nu}^{\alpha} -
\partial^{\alpha}\partial_{\nu}\Phi^{\nu}_{\beta} +
\partial^{\alpha}\partial_{\beta}\Phi_{\nu}^{\nu} -
\delta^{\alpha}_{\beta}\Box\Phi^{\nu}_{\nu} +  \nonumber \\
+ \delta^{\alpha}_{\beta}\partial_{\nu}\partial_{\mu}\Phi^{\nu\mu} =
- 8\pi GT^{\alpha}_{\beta} ,  \label{5.4} \\
\Box e^{\alpha}_{\beta} - \partial_{\beta}\partial^{\nu}e_{\nu}^{\alpha} -
\partial^{\alpha}\partial_{\nu}e^{\nu}_{\beta} +
\partial^{\alpha}\partial_{\beta}e_{\nu}^{\nu} -
\delta^{\alpha}_{\beta}\Box e^{\nu}_{\nu} +
\delta^{\alpha}_{\beta}\partial_{\nu}\partial_{\mu}e^{\nu\mu} +
 \nonumber \\ 
+ m^{2}(e^{\alpha}_{\beta} -
\frac{1}{2}\delta^{\alpha}_{\beta}e^{\nu}_{\nu}) =
- \frac{a}{2}T^{\alpha}_{\beta}  .
\label{5.5}
\end{eqnarray}
Here (\ref{5.4}) are Einshtein equations in the Newton's approximation,
(\ref{5.5}) are translation field equations (\ref{4.12}) in the linear
approximation for the symmetrical part (\ref{4.14}), and $a$ is the coupling 
constant of the translation gauge field. 

For the energy-momentum tensor of the point mass $M$ in the rest, i.e.
\begin{equation}
T^{0}_{0} = M\delta (r) ,~~~~~T^{i}_{\mu} = 0 . \label{5.6}
\end{equation}
equations (\ref{5.4}) and (\ref{5.5}) possesses the static solutions
\begin{equation}
\Phi^{0}_{0} = - \frac{GM}{r}, ~~~~~
e^{0}_{0} = - \frac{aM}{4\pi r}e^{-mr}.  \label{5.7}
\end{equation}

From the effective metric tensor
\begin{equation}
\tilde{g}_{00} = 1 + 2(\Phi_{00} + e_{00})
\label{5.8}
\end{equation}
we obtain the modification of the Newtonian gravitational potential
\begin{equation}
\Phi_{00} + e_{00} = - \frac{GM}{r}(1 +
\frac{a}{8\pi G}e^{-mr}) . \label{5.9}
\end{equation}

Such a modification of the Newton's potential (whose experimental 
verification
received much attention in the 80s) is usually related to the hypothetical
"fifth" fundamental force \cite{FSSTA}. This interaction must be described
by massive classical field. The mass expressed by means of the constants 
$\mu$ and $\lambda$, having the sense of coefficients of "elasticity" of a
space-time, and can be unusually small. On the experimental side, existing
laboratory, geophysical and astronomical data (see references in
\cite{FSSTA}) make restrictions on the value of coupling constants of 
the translation gauge field
\begin{equation}
a < G ,~~~~~m < 10^{-8} ev  ,
\label{5.10}
\end{equation}
where $G$ is the gravitational constant. 

Note that the potential (\ref{5.8}) with parameters $a/8\pi G \sim - 1$ and
$1/m \sim 10 kpc$ may contribute to the problem of mass discrepancies in
galaxies \cite{Sa}. 

Another effect of the massive translation gauge field in the Newtons
approximation can be the difference of the fields of a ball and sphere of
the same mass. 

For the ball of the uniform density
\begin{equation}
T = T^{0}_{0} = \rho ,
\label{5.11}
\end{equation}
with the radius $R$, one can obtain by standard way \cite{LL} the potential 
for the outside region  
\begin{equation}
(\Phi^{0}_{0} + e^{0}_{0})_{r > R} = - \frac{4}{3}\pi R^{3}\frac{G\rho}
{r}\{ 1 + \frac{3ae^{-mr}}{8\pi Gm^{2}R^{3}}[R\cosh (mR) -
\frac{1}{m}\sinh (mR)]\} . \label{5.12}
\end{equation}

For the inside region we have:
\begin{eqnarray}
(\Phi^{0}_{0} + e^{0}_{0})_{r < R} = - 2\pi G\rho
\{ R^{2} - \frac{r^{2}}{3} +
\frac{ae^{-mr}}{4\pi Gm^{2}}[\cosh (mr) -  \nonumber \\ -
\frac{e^{-m(R - r)}}{r}\sinh (mr)(R + \frac{1}{m}) + \sinh (mr)]\} .
\label{5.13}
\end{eqnarray}

The similar expression for the outer region of the sphere has the form:
\begin{equation}
(\Phi^{0}_{0} + e^{0}_{0})^{sph}_{r > R} = - 4\pi R^{2}\frac{G\rho}
{r}[1 + \frac{ae^{-mr}}{4\pi GmR}\sinh (mR)] .
\label{5.14}
\end{equation}

So outer potentials for the uniform density ball and sphere are differ
from each other in the case of the massive translation gauge field.

%%%%%%%%%%%%%%%%%%%%%%%%%%%%%%%%%%%%%%%%%%%%%%%%%%%%%%%%%%%%%%%%%%%%%%%%%%
\newpage
\section{Post Newtonian approximation}
\setcounter{equation}{0}

~~~~~To find post Newtonian approximation of the our theory we must solve
nonlinear equations of the gravitational field and Poincar\'{e} gauge
fields and then divide them in the line with respect of the post Newtonian
parameter $ \sim 10^{-6}$. It is difficult way. We can use the different
method using the fact, that Newton's potential $V = MG/r$ in the Sun
system has the order of the post Newtonian parameter and can be used as
the parameter of division.  In the calculations of the light deflection
and radar echo delay we can restrict ourself by the first order of $V$,
what corresponds to the linear approximation.  Thus in calculations of the
post Newtonian approximation we can use linear equations of translation
gauge fields (\ref{5.5}). We had used this equation when finding
corrections to the Newton potential. In differ of the that case we must
use all and not only zero-zero components of this equations. We shell
consider the spherically symmetrical case when $e_{\mu\nu}$ and
$T^{\mu}_{\nu}$ depends only on $\vec{r}$, and regarding the source as
"pointlike", i.  e. of all the components $T^{\mu}_{\nu}$, assume
different from zero only the component
\begin{equation} 
T^{0}_{0} = -M\delta (r) . \label{6.1}
\end{equation}

To find spherical form of equations (\ref{5.5}) is not easy without using
there nonlinear form. Because we at first find two identities from this
equation, which are not depended on coordinate systems
\begin{eqnarray}
\partial_{\mu}[m^{2}(e^{\mu}_{\nu} -
\frac{1}{2}\delta^{\mu}_{\nu}e^{\alpha}_{\alpha}) +
\frac{a}{2}(T^{\mu}_{\nu} -
\frac{1}{2}\delta^{\mu}_{\nu}T^{\alpha}_{\alpha})] = 0 ,  \nonumber \\
\Box e^{\nu}_{\nu} - \partial_{\alpha}\partial^{\beta}e^{\alpha}_{\beta} +
\frac{m^{2}}{2}e^{\nu}_{\nu} = - \frac{a}{4}T^{\nu}_{\nu} .~~~~~~~~~~
\label{6.2}
\end{eqnarray}

Using this identities and the equation of conservation of the energy-momentum
one can get two equivalent forms of equations (\ref{5.5}):
\begin{eqnarray}
(\Box + m^{2})(e^{\mu}_{\nu} -
\frac{1}{2}\delta^{\mu}_{\nu}e^{\alpha}_{\alpha}) = -
\frac{a}{2}(T^{\mu}_{\nu} -
\frac{1}{2}\delta^{\mu}_{\nu}T^{\alpha}_{\alpha})]
+ \frac{a}{2m^{2}}(\partial^{\mu}\partial_{\nu} -
\frac{1}{2}\delta^{\mu}_{\nu}\Box )T^{\alpha}_{\alpha} ,  \label{6.3} \\
\Box e^{\mu}_{\nu} + \partial^{\mu}\partial_{\nu}e^{\alpha}_{\alpha}
- \partial^{\mu}\partial_{\alpha}e^{\alpha}_{\nu}
- \partial_{\nu}\partial^{\alpha}e^{\mu}_{\alpha} +
m^{2}e^{\mu}_{\nu} = - \frac{a}{2}T^{\mu}_{\nu} .~~~~~~~~~~ \label{6.4}
\end{eqnarray}

From the equation (\ref{6.3}) we noticed that in the spherically
symmetrical case (\ref{6.1}) nondiagonal components of the field
$e^{\mu}_{\nu}$ are zero.  From equation (\ref{6.4}) we see, that
components $e^{\theta}_{\theta}, e^{\phi}_{\phi} $ in this case also are
zero. 

Using the only nonzero components $e^{t}_{t}, e^{r}_{r}$  we get
\begin{eqnarray}
e^{\alpha}_{\alpha} = e^{t}_{t} + e^{r}_{r} ,~~~~~~ \nonumber \\
e^{t}_{t} - \frac{1}{2}e^{\alpha}_{\alpha} = \frac{1}{2}(e^{t}_{t} -
e^{r}_{r}) , \label{6.5} \\
e^{r}_{r} - \frac{1}{2}e^{\alpha}_{\alpha} = - \frac{1}{2}(e^{t}_{t} -
e^{r}_{r}) . \nonumber
\end{eqnarray}
Then zero-zero components of equations (\ref{6.3}) and (\ref{6.4}), 
are:
\begin{eqnarray}
(- \triangle + m^{2})(e^{t}_{t} - e^{r}_{r}) = - \frac{aM}{2m^{2}}
(- \triangle + m^{2})\delta (r) ,   \label{6.6} \\
(- \triangle + m^{2})e^{t}_{t} = - \frac{aM}{2}\delta (r) .~~~~~~~~~~~~~~~ 
\label{6.7} \end{eqnarray}

From (\ref{6.6})  we get:
\begin{equation}
e^{t}_{t} - e^{r}_{r} = - \frac{aM}{8\pi m^{2}}\delta (r) , \label{6.8}
\end{equation}
(this relation can be obtained from (\ref{6.2})  immediately using 
(\ref{6.1}) and (\ref{6.5})), and from (\ref{6.7}) we have:
 \begin{equation}
 e^{t}_{t} = - \frac{aM}{8\pi}\frac{e^{-mr}}{r} .
\label{6.9}
\end{equation}

Finally solution of equations of  translation gauge field in the linear
approximation is:
\begin{eqnarray}
e^{t}_{t} = - \frac{aM}{8\pi}\frac{e^{-mr}}{r} ,~~~~~
e^{r}_{r} =  \frac{aM}{8\pi}[\frac{\delta (r)}{m^{2}} - \frac{e^{-mr}}{r}] ,
 \label{6.10} \\
e^{\theta}_{\theta} = e^{\phi}_{\phi} = 0 , ~~~~~e^{\mu}_{\nu} = 0 
~~~~~(\mu \neq \nu). \nonumber
\end{eqnarray}

Using solutions (\ref{6.10}) and solution of the Einshtein equations in 
linear approximation (\ref{5.3}) - 
\begin{eqnarray}
\Phi^{t}_{t} = - V = - \frac{GM}{r} ,~~~~~
\Phi^{r}_{r} = V = \frac{GM}{r}  \nonumber \\
\Phi^{\theta}_{\theta} = \Phi^{\phi}_{\phi} = 0 , ~~~~~\Phi^{\mu}_{\nu} = 0 
~~~~~(\mu \neq \nu) ,  \label{6.11}
\end{eqnarray}
we can belt the effective metric tensor of the theory in the linear 
approximation \begin{equation}
\tilde{g}_{\mu\nu} = \eta_{\mu\nu} + 2(e_{\mu\nu} + \Phi_{\mu\nu}) ,
\label{6.12}
\end{equation}
in the form:
\begin{eqnarray}
\tilde{g}_{tt} = 1 - 2V(1 + \frac{a}{8\pi G}e^{-mr}) ,  \nonumber \\
\tilde{g}_{rr} = -1 - 2V(1 - \frac{a}{8\pi G}e^{-mr}) , \label{6.13}\\ 
\tilde{g}_{\theta\theta} = - r^{2} ,~~~~~
\tilde{g}_{\phi\phi} = - r^{2}\sin ^{2}\theta . \nonumber
\end{eqnarray}

Equation for particles moving on geodesic lines has the form:
\begin{equation}
\frac{d^{2}x^{\mu}}{\tilde{d}s^{2}} +
\tilde{\Gamma}^{\mu}_{\alpha\beta}\frac{dx^{\alpha}}{\tilde{d}s}
\frac{dx^{\beta}}{\tilde{d}s} = 0 . \label{6.14}
\end{equation}
Here $\tilde{\Gamma}^{\mu}_{\alpha\beta} $ are effective Christoffel
symbols builded by effective metric (\ref{6.13}). Then by standard way
\cite{We} we can get formulae for the light deflection angle
\begin{equation}
\phi = \int^{\infty}_{r'}(-\tilde{g}_{rr})^{1/2}[ \frac{\tilde{g}'_{tt}}
{r'^{2}}\tilde{g}_{tt} - \frac{1}{r^{2}}] ^{-1/2}\frac{dr}{r^{2}} ,
\label{6.15}
\end{equation}
and for the radar echo delay time
\begin{equation}
t = \int^{r_{2}}_{r_{1}}(-\tilde{g}_{rr}\tilde{g}'_{tt})^{1/2}
[ \frac{\tilde{g}'_{tt}}{r'^{2}}\tilde{g}_{tt} - \frac{1}{r^{2}}] ^{-1/2}
\frac{dr}{r'\tilde{g}_{tt}} , \label{6.16}
\end{equation}
in fields of the spherical mass.  Putting the solution (\ref{6.13}) in this 
equations, in the linear approximation by the Newtons potential we obtain
\begin{eqnarray}
\phi = (\frac{\pi}{2} + 2\frac{MG}{r'}) + \frac{aM}{8\pi}[\frac{\pi}{2} -
\int^{mr'}_{0}K_{0}(mr')d(mr')] ,  \nonumber \\
t = (\sqrt{r^{2}_{2} - r^{2}_{1}} + MG\sqrt{\frac{r_{2} - r_{1}}
{r_{2} + r_{1}}} + 2MG\ln \mid\frac{r_{2} + \sqrt{r^{2}_{2} - r^{2}_{1}}}
{r_{1}}\mid) - \label{6.17}\\ 
- \frac{aMr^{2}_{1}}{8\pi}\{\frac{e^{-mr_{1}}}{
\sqrt{r^{2}_{2} - r^{2}_{1}}} + m[K_{0}(mr_{2}) - K_{0}(mr_{1})]\} ,
\nonumber
\end{eqnarray}
where $K_{0}(x) = \lim_{\nu \rightarrow 0} K_{\nu}(x)$  is a MacDonald's
function of the zero order. First terms in this expressions are values of in
the Einshtein theory and second terms are corrections by the translation 
gauge field.

%%%%%%%%%%%%%%%%%%%%%%%%%%%%%%%%%%%%%%%%%%%%%%%%%%%%%%%%%%%%%%%%%%%%%%%%
\newpage
\section{The string like solution}
\setcounter{equation}{0}

~~~~~The analogy with the linear defect in crystals leads to the search of
the string like solution of the theory. Two-dimensional objects, for example
cosmic strings, are considered in field theories. They are  solutions
of the hypothetical charged scalar field equations and suggested to appear in
early stages of the universe expansion. Due to specific form of the string 
core energy-momentum tensor
\begin{equation}
T^{t}_{t} = T^{z}_{z} = \frac{1}{2}T^{\nu}_{\nu} ,
\label{7.1}
\end{equation}
strings does not act gravitationally with the surrounding matter. For the
cosmic string the solution of Einshtein equations in cylindrical
coordinates are:
\begin{equation}
ds^{2} = (dt - \alpha d\theta)^{2} - dr^{2} - r^{2}b^{2}d\theta^{2} - dz^{2} .
\label{7.2}
\end{equation}
This solution globally corresponds to the cone, but is locally flat 
\cite{Vi}.

It can be checked, that in the vacuum ($T_{\mu\nu} = 0$) nonlinear 
translation gauge field equations (\ref{4.12}) have the solution
\begin{eqnarray}
\theta^{t}_{x} = \theta^{z}_{x} = \frac{\alpha y}{x^{2} + y^{2}} , 
\nonumber \\
\theta^{t}_{y} = \theta^{z}_{y} = - \frac{\alpha x}{x^{2} + y^{2}} ,
\label{7.3}
\end{eqnarray}
where $\alpha$ is arbitrary constant. Indeed, in this case components of
the torsion tensor (\ref{4.2}) and, thus the Lagrangian (\ref{4.6}) are 
zero. Nonzero components of (\ref{4.5}) are:  
\begin{eqnarray}
e_{xz} = - e_{xt} = - \frac{\alpha y}{x^{2} + y^{2}} ,  \nonumber \\
e_{yz} = - e_{yt} =  \frac{\alpha x}{x^{2} + y^{2}} ,
\label{7.4}
\end{eqnarray}
Thus the Lagrangian (\ref{4.4}), and combinations $e^{\nu}_{\nu}$ and
$e_{\gamma\mu} h^{\gamma}_{\nu}$ are also zero, and equations
(\ref{4.12}) are satisfied identically. 

The components of the effective metric tensor (\ref{3.14}), without 
inserting 
there gravity, for the solution (\ref{7.3}) in cylindrical coordinates has 
the form: \begin{eqnarray}
\tilde{g}_{tt} = - \tilde{g}_{rr} = - \tilde{g}_{zz} = 1 , \nonumber \\
\tilde{g}_{\theta\theta} = - r^{2} ,~~~~~~~~~~~~ \label{7.5} \\
\tilde{g}_{\theta z} = - \tilde{g}_{\theta t} = \alpha .~~~~~~ \nonumber
\end{eqnarray}

The reverse tensor has  components:
\begin{eqnarray}
\tilde{g}^{tt} = 1 - \frac{\alpha^{2}}{r^{2}},~~~~~ \tilde{g}^{rr} = - 1 ,
\nonumber \\
\tilde{g}^{\theta\theta} = - \frac{1}{r^{2}} ,~~~~~
\tilde{g}^{zz} = -1 - \frac{\alpha^{2}}{r^{2}} , \label{7.6} \\
\tilde{g}^{\theta z} =  \tilde{g}^{\theta t} = - \frac{\alpha}{r^{2}} ,
~~~~~\tilde{g}^{tz} = - \frac{\alpha^{2}}{r^{2}} .
\nonumber
\end{eqnarray}

Calculation of the energy-momentum tensor and effective interval for the
 solution (\ref{7.3}) gives:
\begin{eqnarray}
T_{tt} = - T_{zz} = T_{zt} = \frac{\alpha^{2}}{r^{2}} ,~~~~~~~~~~~~~~~ 
\nonumber \\ \tilde{d}s^{2} = (dt - \frac{\alpha}{2} d\theta)^{2} -
dr^{2} - r^{2}d\theta^{2} - (dz + \frac{\alpha}{2}d\theta)^{2} .
\label{7.7}
\end{eqnarray}

We see that this result is similar south for the cosmic string 
(\ref{7.1}) and
(\ref{7.2}). It means that solution (\ref{7.3}) corresponds to the string 
like object in the space-time. 

Now let's consider motion of particles in the  string like  solution
(\ref{7.3}). The scalar field $\psi$ 
motion equation in the external translation gauge field, expressed by 
the effective metric tensor (\ref{7.6}) is: 
\begin{equation}
(\tilde{g}^{\mu\nu}\partial_{\mu}\partial_{\nu} + \mu^{2})\psi (x) = 0 .
\label{7.8}
\end{equation}

We search the solution in the form:
\begin{equation}
\psi (t,r,\theta,z) = \int dkd\omega e^{i(kz + \omega t)}
\sum^{+ \infty}_{m = - \infty}e^{im\theta}\varphi (r) .
\label{7.9}
\end{equation}

For the radial wave function we get the Bessel equation
\begin{equation}
\{\frac{\partial^{2}}{\partial r^{2}} + \frac{1}{r}
\frac{\partial}{\partial r} - \frac{[m + \alpha (k + \omega)]^{2}}{r^{2}}
+ P^{2}\}\varphi (r) = 0 ,
\label{7.10}
\end{equation}
where $P^{2} = \omega^{2} - k^{2} - \mu^{2}$. This is a Aharonov-Bohm
equation, having the solution:
\begin{equation}
\varphi (r) \longrightarrow e^{- i(\alpha\theta + Pr\cos \theta)} \pm
\frac{e^{iPr}}{\sqrt{2\pi iPr}}\sin (\pi\alpha)
\frac{e^{\mp i\theta /2}}{\cos (\theta /2)} ,
\label{7.11}
\end{equation}

In this expression signs $\pm$ correspond to particles going around the
string from different sides. We get analogy of the Aharonov-Bohm effect:
in spite of absence of the string field strength outside the core,
particles feel this field (the interference pattern changes behind the
string). Here the effect exists even for spinless particles. 

Another possible application of the string like translation gauge field
can be galaxy formation. So far in the cosmology existed two mechanisms of
generation of density fluctuations on which afterwards were formed
galaxies. First of them basing on a inflation scenario connects density
fluctuations with quantum fluctuations of the Higgs field in the early
universe. Such approach requires the introduction of a dark matter of
vague nature. Alternatively the second mechanism considers galaxy
formation on strings loops, might be born during phase transition in the
early universe and were evaporated due to the gravitational radiation
\cite{Vi}. But in this approach must be allowed existence of the some
scalar field forming the cosmic string. The translation gauge field can
serve as the new approach which leads to the stable density fluctuations
in the early universe. In the difference of other models the translation
gauge field appears in the theory from fundamental principles and does not
require the introduction of supplementary fields. 

%%%%%%%%%%%%%%%%%%%%%%%%%%%%%%%%%%%%%%%%%%%%%%%%%%%%%%%%%%%%%%%%%%%%%%%%%%%
\newpage
\section{Spherically symmetrical solution}
\setcounter{equation}{0}

~~~~~To the end of this paper we want to find a spherical-symmetrical
solution of nonlinear system of equations of the gravitational field and
translation gauge field (\ref{4.12}) in outer region of the spherically 
symmetrical source, i.e.  for the case 
\begin{equation} T^{\mu}_{\nu} = 0 .
\label{8.1} \end{equation}

In this section we consider that Lorentz gauge fields are zero -
$W^{A}_{\mu} = 0$, and the translation gauge field has not mass, so 
constants $\mu$ and $\lambda$ also are zero.  

In the equations of translation gauge fields (\ref{4.12}) we put 
\begin{eqnarray}
\theta^{0}_{0} = A(r) ,~~~~~\theta^{i}_{j} = \delta^{i}_{j} B(r) ,
\nonumber \\
\theta^{0}_{i} = \theta^{i}_{0} = 0 ~~~~~(i,j = 1,2,3) .
\label{8.2}
\end{eqnarray}
From the formula (\ref{4.2}) we can find all nonzero components of field
strength
\begin{eqnarray}
S^{0}_{xo} = \partial _{x}A , ~~~~~S^{y}_{xy} = F^{z}_{xz} = \partial_{x}B ,
 \nonumber \\
S^{0}_{yo} = \partial _{y}A , ~~~~~S^{x}_{yx} = F^{z}_{yz} = \partial_{y}B ,
\label{8.3} \\
S^{0}_{zo} = \partial _{z}A , ~~~~~S^{x}_{zx} = F^{y}_{zy} = \partial_{z}B .
\nonumber
\end{eqnarray}

So equations (\ref{4.12}) takes the form:
\begin{eqnarray}
2g_{00}\triangle B = 0 ,~~~~~~~~~~~~~~~~~~~~  \nonumber \\
g_{xx}[\triangle (A + 2B) - \partial^{2}_{x}(A + 2B) -
(\partial^{2}_{y} + \partial^{2}_{z})B] = 0 ,  \nonumber \\
g_{yy}[\triangle (A + 2B) - \partial^{2}_{y}(A + 2B) -
(\partial^{2}_{x} + \partial^{2}_{z})B] = 0 ,  \nonumber \\
g_{zz}[\triangle (A + 2B) - \partial^{2}_{z}(A + 2B) -
(\partial^{2}_{x} + \partial^{2}_{z})B] = 0 , \label{8.4} \\
\partial_{x}\partial_{y}(A + B) = 0 ,~~~~~~~~~~~~~~~  \nonumber \\
\partial_{x}\partial_{z}(A + B) = 0 ,~~~~~~~~~~~~~~~  \nonumber \\
\partial_{y}\partial_{z}(A + B) = 0 .~~~~~~~~~~~~~~~  \nonumber
\end{eqnarray}
From the latter three equations we have:
\begin{equation}
A(r) = - B(r) .
\label{8.5}
\end{equation}
Then all of equations (\ref{8.4})  are satisfied but first, which gets the 
form: \begin{equation}
\triangle A(r) = 0 .
\label{8.6}
\end{equation}
Thus the spherical-symmetrical solution of translation gauge field 
equations (\ref{4.12}) is:
\begin{eqnarray}
\theta^{0}_{0} = \frac{c}{r} ,~~~~~~~~~~
\theta^{i}_{j} = - \delta^{i}_{j}\frac{c}{r} ,  \nonumber \\
\theta^{i}_{0} = \theta^{0}_{i} = 0, ~~~~~i,j = 1,2,3.
\label{8.7}
\end{eqnarray}
Here $c$ is intagrable constant which must be fixed by boundary conditions.
In spherical coordinates solution (\ref{8.7}) has the form:
\begin{eqnarray}
\theta^{0}_{0} = \frac{c}{r} ,~~~~~\theta^{r}_{r} = - \frac{c}{r} ,
 \nonumber \\
\theta^{\theta}_{\theta} = \theta^{\phi}_{\phi} = 0 .~~~~~~~~~~
\label{8.8}
\end{eqnarray}

Now let's remain the spherical - symmetrical solution of Einshteins
 equations - the Schwarzschild solution
\begin{eqnarray}
g_{00} = - g^{-1}_{rr} = 1 + \frac{E}{r} , ~~~~~~~~~~ \nonumber \\
g_{\theta\theta} = - r^{2},~~~~~g_{\phi\phi} = - r^{2}\sin^{2}\theta ,
\label{8.9}
\end{eqnarray}
where $E$ also is a intagrable constant.

 Using the definition of the effective metric (\ref{3.14})   we have:
\begin{eqnarray}
\tilde{g}_{00} = (1 + \frac{E}{r})(1 + \frac{c}{r})^{2} , \nonumber \\
\tilde{g}_{rr} = - (1 + \frac{E}{r})^{-1}(1 - \frac{c}{r})^{2} , 
\label{8.10} \\
\tilde{g}_{\theta\theta} = - r^{2},~~~~~
\tilde{g}_{\phi\phi} = - r^{2}\sin^{2}\theta . \nonumber
\end{eqnarray}

 Values for constants $c, E$  in this expressions can be obtained from
the Newton's approximation. For the massless translation gauge field in 
the  Newton's approximation we have: 
\begin{equation}
\tilde{g}_{00} \simeq 1 + 2\frac{M(G - a)}{r} ,
\label{8.11}
\end{equation}
 where $M$ is mass of the source, $G$ is gravitational constant and $a$ is
the gauge field constant. So we have
\begin{equation}
E= - 2MG , ~~~~~c = 2Ma .
\label{8.12}
\end{equation}

 Finally the spherical-symmetrical solution in the tensor-tensor theory of
 gravity has the following form:
\begin{eqnarray}
\tilde{g}_{00} = (1 - \frac{2MG}{r})(1 + \frac{2Ma}{r})^{2} ,  \nonumber \\
\tilde{g}_{rr} = - (1 - \frac{2MG}{r})^{-1}(1 - \frac{2Ma}{r})^{2} , 
\label {8.13} \\  
\tilde{g}_{\theta\theta} = - r^{2},~~~~~
\tilde{g}_{\phi\phi} = - r^{2}\sin^{2}\theta . \nonumber
\end{eqnarray}

At the end of this section we want to consider a gravitational 
collapse problem. From the solution (\ref{8.13}) one noticed, that, 
since we have relations (\ref{5.10}), inside the Schwarzschild sphere
\begin{equation}
r = 2MG ,
\label{8.14}
\end{equation}
which is half-penetrable for the matter to the singular point
\begin{equation}
r = 0 ,
\label{8.15}
\end{equation}
there exists second, the real singular surface
\begin{equation}
r = 2Ma ,
\label{8.16}
\end{equation}
which is impenetrable for the matter. Indeed equations of the translation
gauge field don't let Kruskal like coordinate transformations, because of
the translation gauge field is single valued and it can't be removed from
the effective metric by coordinate transformations. Also determinant of the
gauge field
\begin{equation}
det(\delta^{\mu}_{\nu} + \theta^{\mu}_{\nu}) =
(1 + \frac{2Ma}{r})(1 - \frac{2Ma}{r})
\label{8.17}
\end{equation}
on the sphere (\ref{8.16})  is singular. So existence of the translation 
gauge field avoid the collapse problem.

\vskip 0.5cm
This work was supported in part by International Science Foundation (ISF)
under the Grant No MXL000.
%%%%%%%%%%%%%%%%%%%%%%%%%%%%%%%%%%%%%%%%%%%%%%%%%%%%%%%%%%%%%%%%%%%%%

\end{document}